# Giant Orbital Torque-driven Picosecond Switching in Magnetic Tunnel Junctions

**Authors:** Yuxuan Yao[1]†, Chen Xiao[1]†, Xiaobai Ning[1,4]†, Wenlong Cai[1]†, Xianzeng Guo[1], Zongxia Guo[1,5], Kailin Yang[1], Danrong Xiong[3], Zhengjie Yan[1], Shiyang Lu[1], Hongchao Zhang[3], Siyuan Cheng[1], Renyou Xu[1], Dinghao Ma[1], Chao Wang[1,2], Zhaohao Wang[1,2], Daoqian Zhu[1,2]*, Kaihua Cao[1], Hongxi Liu[3]*, Aurélien Manchon[4] and Weisheng Zhao[1,2]*

**Affiliations:**

[1]Fert Beijing Institute, School of Integrated Circuit Science and Engineering, Beihang University; Beijing, 100191, China.

[2]National Key Lab of Spintronics, Institute of International Innovation, Beihang University; Hangzhou, 311115, China.

[3]Truth Memory Corporation; Beijing, 100088, China.

[4]CINaM, Aix-Marseille Univ, CNRS; Marseille, France.

[5]Laboratoire Albert Fert, CNRS, Thales, Université Paris-Saclay; Palaiseau, 91767, France.

†These authors contributed equally to this work.

* Corresponding author. Email: weisheng.zhao@buaa.edu.cn; daoqian_zhu@buaa.edu.cn; hongxi_liu@tmc-bj.cn

**Abstract:** Orbital Hall effect was recently discovered as a novel pathway for driving magnetic moment. However, the integration of orbital Hall effect in magnetic memories suffers from low orbital-to-spin conversion efficiency and incompatibility with magnetic tunnel junctions. Here we demonstrate an orbital Hall effect-driven magnetic tunnel junction based on Ru/W bilayer, where the Ru layer possesses a strong orbital Hall conductivity and the $\alpha$-W layer features an orbital-to-spin conversion efficiency exceeding 90% because of the large orbit-spin diffusivity. By harnessing the giant orbital torque, we achieve a 28.7-picosecond switching and a five to eight-fold reduction in driving voltages over conventional spin-orbit torque magnetic memories. Our work bridges the critical gap between orbital effects and magnetic memory applications, significantly advancing the field of spintronics and orbitronics.

**Main Text:**

The efficient generation of spin currents is a cornerstone of spintronics, particularly in the advancement of magnetic random-access memory (MRAM) for high-performance memory and logic applications(*1–6*). In spin-transfer torque MRAM, spin currents are generated by a ferromagnetic (FM) electrode in magnetic tunnel junctions (MTJs) with a core structure of FM/tunneling barrier/FM(*7, 8*). However, the necessity for write currents to traverse the tunneling barrier significantly limits the write speed and device endurance. In contrast, spin-orbit torque MRAM (SOT-MRAM) leverages charge-to-spin conversion in an adjacent non-magnetic layer(*9–11*). This architecture enables spin current generation with comparable efficiency to that of FM electrodes, while the write currents bypass the tunneling barrier. Thus, SOT-MRAM offers enhanced speed and durability, emerging as a highly promising candidate for next-generation memory and logic applications.

To date, the non-magnetic SOT write channels predominantly employ materials exhibiting strong spin-orbit coupling (SOC), such as *β*-W(*12, 13*) and topological insulators(*14, 15*), where the spin currents are directly converted from charge currents via spin Hall effect (SHE) or Rashba-Edelstein effect. While materials with a satisfying charge-to-spin conversion efficiency ($\theta_{SH} > 0.3$) are typically situated in the dirty regime with resistivities of 200-1000 $\mu\Omega\cdot$cm, their spin Hall conductivity (SHC) remains limited, often below that of Pt(*16–19*), approximately 4400 $\hbar/2e$ $\Omega^{-1}$ cm$^{-1}$. The large and non-uniform resistance of the bottom electrode (BE), which serves as the write channel in SOT devices, leads to excessively high writing voltages that may exceed the driving capability of transistors. Moreover, this resistance variability contributes to significant device-to-device performance distribution. This inherent trade-off poses severe challenges for the continuous scaling of SOT-MRAM, as well as for enhancing writing speed and device uniformity.

The recent discovery of orbital Hall effect (OHE) and orbital Rashba-Edelstein effect (OREE) provides a promising path to address the aforementioned issues(*20–24*). These orbital effects enable the generation of orbital angular momentum (OAM) in response to an applied electric field, even in light metals of weak SOC and low resistivity (Ru(*25, 26*), CuO$_x$(*22, 27, 28*)). The arisen OAM provides an individual degree of freedom to generate spin currents through orbital-to-spin conversion, thus presenting new opportunities for SOT-MRAM. Though theoretical predictions suggest that the orbital Hall conductivity (OHC) is substantially stronger than its spin counterpart(*16, 29*), experimental measurements of OHC have fallen short of the values required for high performance magnetic memories(*20, 21, 23, 30*). This discrepancy exists because the local spin moment of FM storage layer cannot directly interact with the OAM *L*, but the conversion efficiency $C_{LS}$ from *L* to spin angular momentum (SAM) *S* in the converting layer was found to be inadequate(*23*). Therefore, the measured orbital torque (OT) and effective spin-orbit Hall conductivity $\sigma_{eff}$ are below expectations. Another major challenge lies in the incompatibility between OT channels and MTJs. Materials with high $C_{LS}$, such as Ni(*23, 31, 32*) or Gd(*21, 30*), are typically difficult to integrate with CoFeB/MgO/CoFeB core layers, which are essential for achieving large perpendicular magnetic anisotropy (PMA) and tunneling magnetoresistance (TMR) required for MRAM scaling and readout(*8, 33*). Here, we address these critical challenges by demonstrating OHE-MTJs with a Ru/W-based OT channel (See Movie S1). We obtain a giant $\sigma_{eff}$ of -12,631 $\hbar/2e$ $\Omega^{-1}$ cm$^{-1}$, corresponding to a $C_{LS}$ of 94% that is attributed to the orbit-spin diffusivity significantly higher than the orbital diffusivity(*34*). Leveraging the giant OT and the low resistivity of the Ru/W bilayer (20 $\mu\Omega\cdot$cm), we achieve a 28.7-picosecond switching of the OHE-MTJs. Furthermore, through circuit-level simulations, we demonstrate a 32-bit NAND-like OHE-MTJ array that reduces device area by 45% compared to *β*-W-based counterparts. These advancements highlight the potential of OHE-MTJs to significantly enhance the performance and scalability of next-generation MRAM technology.

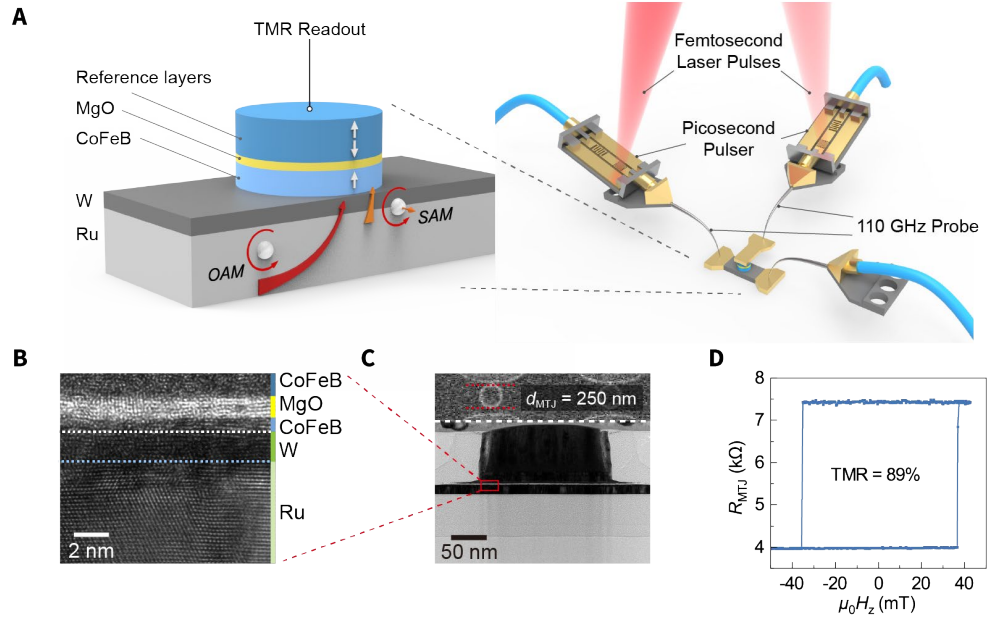

**Fig. 1. Schematic and device characterization of the OHE-MTJ.** (**A**) Schematic of the OHE-MTJ device and the experimental setup for picosecond switching. Orbital current is generated within the Ru layer, converted to spin current within the W layer and consequently switches the CoFeB free layer. Picosecond electrical pulses are generated by two standalone photoconductive switches, then applied on the OT channel through two 110 GHz probes. (**B**) Cross-sectional TEM image of an OHE-MTJ device, highlighting distinct layer interfaces delineated by dashed lines. (**C**), Combined scanning electron microscopy and TEM image of an OHE-MTJ device, indicating the device diameter $d$ of 250 nm. (**D**)*R-H* hysteresis loop of the OHE-MTJ device, demonstrating a TMR ratio of 89%.

**Device geometry and magnetic performance**

The schematic of the proposed OHE-MTJs is illustrated in Fig. 1**A**. The OT channel is constructed using a Ru/W bilayer, where Ru and W serve as the orbital source layer and *L*-to-*S* converting layer, respectively. After the *L*-to-*S* conversion, the magnetization of CoFeB free layer is switched by the generated OT, and can be readout by measuring the TMR of the device. Fig. 1A also includes a simplified schematic of the experimental set-up for OT-induced picosecond switching, as discussed in detail in subsequent sections.

The stack of the OHE-MTJ, composing of substrate/Ta(0.5)/Ru(12) /W(1.6)/CoFeB(0.8)/ MgO(1.2)/CoFeB reference layers/W/[Co/Pt]$_m$/Co/Ru/[Co/Pt]$_n$/ Ru/top electrodes (thickness in nanometers), was prepared by magnetron sputtering at room temperature, followed by a post-annealing process at 350°C. The CoFeB-based free layer and reference layer maintain robust PMA after the annealing process, as shown in fig. S4A. Fig. 1**B** reveals a well-textured hcp Ru layer, bcc *α*-W, and a (001)-textured CoFeB/MgO/CoFeB sandwiched structure that guarantees a large TMR ratio (see Supplementary Note S2 for detailed analysis). Subsequently, we fabricated the full stack into OHE-MTJ devices with a diameter $d = 250$ nm, as shown in Fig. 1C. The BE of OHE-MTJs features a length of 1150 nm and a width of 420 nm. Fig. 1D shows the $R$-$H$ hysteresis loop, exhibiting a large TMR ratio of 89%.

In OHE-MTJs, Ru is selected as the orbital source because our first-principles calculations(35, 36) reveal an OHC of 13402 $\hbar/2e$ $\Omega^{-1}$ cm$^{-1}$ (or even higher in ref.(16)) and a SHC of 182 $\hbar/2e$ $\Omega^{-1}$ cm$^{-1}$ at the Fermi level for (001)-oriented hcp Ru (Fig. 2A). This substantial disparity between the OHC and SHC is advantageous for discerning the switching mechanism, while the robust OHC facilitates Ru as a highly promising candidate for the OT channel. Additionally, our calculations based on tight-binding approaches(34) demonstrate that the $C_{LS}$ can be as high as -0.75 for $\alpha$-phase crystalline W, as depicted in Fig. 2B, consistent with cross-sectional transmission electron microscopy (TEM) results presented in Fig. 1B. Notably, this highly efficient $C_{LS}$ at the Fermi level originates from the higher orbit-spin diffusivity relative to the orbital diffusivity. The sign fluctuation of $C_{LS}$ with respect to the chemical potential is examined by investigating the spin-orbit correlation shown in Fig. 2C. The band structure in the vicinity of H exhibits a negative spin-orbit correlation, which determines the negative sign of the $C_{LS}$ within a narrow energy range around the Fermi level.

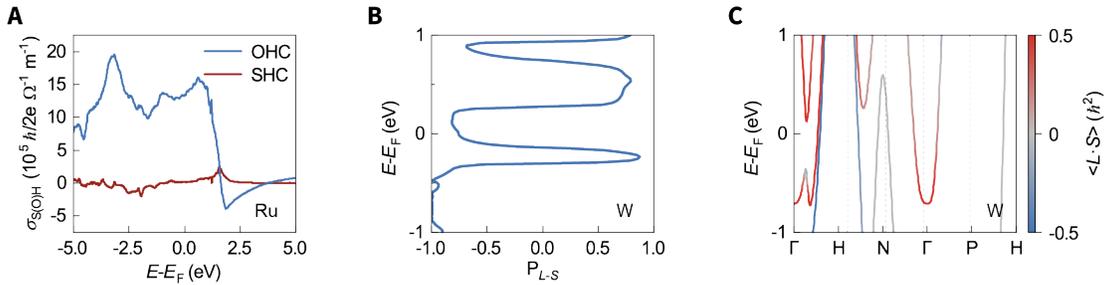

**Fig. 2. First-principles calculations.** (**A**) Calculated spin Hall conductivity and orbital Hall conductivity of Ru. (**B**) Calculated orbit-spin polarization efficiency $P_{L\text{-}S}$, i.e., orbital-to-spin conversion efficiency $C_{LS}$, of $\alpha$-W using tight-binding approaches. (**C**) Spin-orbit correlation $\langle L \cdot S \rangle$ projected onto the band structure of $\alpha$-W.

## Giant OT torque generated with Ru/W channel

We then measured the effective charge-to-spin conversion efficiency $\theta_{\text{eff}}$ and effective spin-orbit Hall conductivity $\sigma_{\text{eff}}$ in Ru/W-based OT channels. We prepared two series of stacks, namely, Ta(0.5)/Ru($t_{\text{Ru}}$)/W(1.5)/CoFeB(0.8)/MgO(1.6)/Ta(1.5) and Ta(0.5)/Ru(12)/W($t_{\text{W}}$)/CoFeB(0.8)/MgO(1.6)/Ta(1.5), where the thicknesses of Ru and W layers are systematically varied to elucidate the OHE contribution. These stacks were patterned into Hall bar devices with a width of 3 μm, as shown in the inset of Fig. 3A. To characterize the $\theta_{\text{eff}}$, we performed second harmonic Hall voltage measurements for the two series of samples. By applying an ac current along the $x$ axis, second harmonic Hall signal $R_{2\omega}$ was measured along the $y$ axis by a lock-in amplifier. By sweeping an external magnetic field $H_x$, larger than the anisotropy field $H_K$, along $x$ axis, the second harmonic resistance $R_{2\omega}$ can be described by(37, 38) $R_{2\omega} = \frac{R_{\text{AHE}}}{2}\frac{H_{\text{DL}}}{|H_x|-H_K} + R_{\text{PHE}}\frac{H_{\text{FL}}}{|H_x|} + R_T + R_0$, where $R_{\text{AHE}}$, $R_{\text{PHE}}$, $R_T$ and $R_0$ are anomalous Hall resistance, planar Hall resistance, resistance contributed by Nernst effects and offset resistance, respectively. Given that the $R_{\text{PHE}}$ is much smaller than $R_{\text{AHE}}$, the second term in the equation was neglected. Thus, we extract the damping-like torque effective field $H_{\text{DL}}$ by excluding thermal contributions and averaging the values from positive and negative field sweeps (see fig. S1A). We further varied the amplitude of the applied ac current, then obtained the $H_{DL}$ through linear fittings, as depicted in fig. S1B. Subsequently, the $\theta_{eff}$ and $\sigma_{eff}$ for both series of samples were calculated(*37, 38*) by $\theta_{DL} = \frac{\mu_0 H_{DL}}{J} M_S t_F \frac{2e}{\hbar}$ and $\sigma_{DL} = \frac{\theta_{DL}}{\rho_{xx}}$. Fig. 3A presents the thickness dependent of $\theta_{eff}$ for samples with varying $t_{Ru}$ by blue triangles. Thickening the Ru layer firstly decreases the $\theta_{eff}$ and then continuously increases the $\theta_{eff}$.

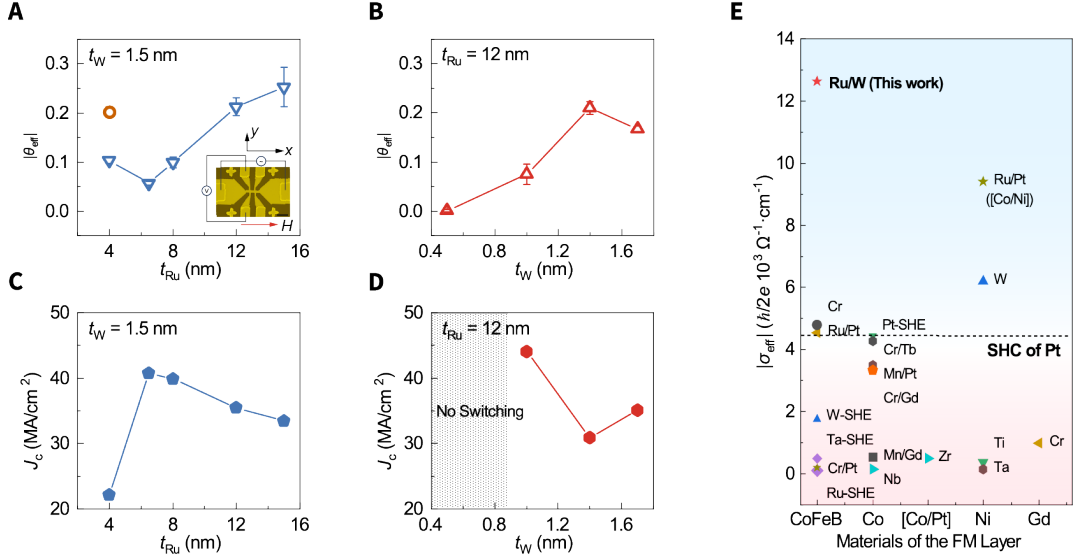

**Fig. 3. Giant OT and strong effective spin-orbit Hall conductivity generated with Ru/W channel.** (**A, B**) The absolute value of effective charge-to-spin conversion efficiency $\theta_{eff}$ in the Ru/W OT channel with varying thickness of (A) Ru and (B) W layers. The thickness of the W and Ru layers is fixed at (A) 1.5 nm and (B) 12 nm, respectively. The orange circle denotes an assumed situation that the $\theta_{eff}$ is solely attributed to the W converting layer. Schematic of second harmonic Hall voltage measurement is shown in the inset of (A). (**C, D**) Threshold write current density $J_c$ of Hall-bar devices corresponding to (A) and (B), respectively. (**E**) Comparison of effective spin-orbit Hall conductivity $\sigma_{eff}$ in Ru/W OT channel against other reported OT or SOT channels. The horizontal axis represents material choices as the free layer in MTJs, with the compatibility of these materials with MTJs decreasing from left to right.

For the device with $t_{Ru}$ = 4 nm, the resistivity $\rho_{xx}$ of the Ru layer is notably high (see fig. S6A), thus the current density applied in the W layer $J_W$ remains at approximately half of that in the Ru layer ($J_{Ru}$), as shown in fig. S6B. If only the W layer was taken into account, namely, assuming that no orbital current is generated in the Ru layer, the effective $\theta_{eff, W}$ is estimated to -0.21 for $t_W$ = 1.5 nm, indicated by an orange circle in Fig. 3A. The assumed $\theta_{eff, W}$ is reasonably large considering the frequently reported large $\theta_{SH}$ in W(*12, 13*). By increasing $t_{Ru}$, the $\rho_{xx}$ of the Ru layer significantly decreases to as low as 20 μΩ·cm, leaving a negligible current shunting through the W layer. For instance, $J_W$ is estimated to be 0.11×$J_{Ru}$ at $t_{Ru}$ = 12 nm. This observation allows us to rule out any predominant contribution of spin currents from the W converting layer. Moreover, the low resistivity of Ru and the TEM image in Fig. 1B collectively demonstrate excellent crystallinity of the Ru layer, excluding potential extrinsic contributions. Fig. 3**B** further shows the $\theta_{eff}$ of the Ru(12)/W($t_W$) stacks. Consistent with previous reports of other bilayer OT channels(*21*), the $\theta_{eff}$ initially increases to saturation, then decreases when the converting layer W is as thick as 1.7 nm. This behavior is attributed to the maximization of orbital-to-spin conversion within the converting layer of specific thickness due to an orbital diffusion length of several nanometers(*21*). In this stack, again, the negligible contribution of spin currents from W is further confirmed by the limited current shunting. The variation of $\theta_{eff}$ with $t_W$ provides additional evidence supporting the dominant contribution of the OHE. Thereafter, we performed current-induced magnetization switching in the aforementioned Hall bar devices (fig. S7). Figs. 3**C** and 3**D** summarize the critical switching current density $J_c$, which shares consistent trends as the second harmonic Hall measurements in Figs. 3A and 3B, further validating the dominant contribution of the OT.

Remarkably, we demonstrate that the $\theta_{eff}$ of the Ru(15)/W(1.5) OT channel reaches -0.25, while the spin-orbit Hall conductivity $\sigma_{eff}$ is as high as -12,631 $\hbar/2e$ $\Omega^{-1}$ cm$^{-1}$. This $\sigma_{eff}$ represents the highest value reported to date (Fig. 3**E**), surpassing the SHC of Pt by a factor of approximately three, and aligns closely with first-principles calculations. Considering the calculated OHC of Ru, we estimate that the conversion efficiency exceeds 90%, corroborating the value computed in $\alpha$-W both in magnitude and sign (see Fig. 2B). Beyond the giant OT, the Ru/W OT channel exhibits excellent compatibility with the CoFeB-based MTJ systems. While materials such as Co/Pt, Ni or rare-earth metals have demonstrated relatively efficient orbital-to-spin conversion, their integration with MTJs is hindered by challenges such as lattice mismatch or inevitable element diffusion, especially after high-temperature post-annealing. In this regard, our Ru/W channel outperforms other reported OT or SOT channels (Fig. 3E)(*20, 21*)·(*23, 25–27, 30, 39–42*). By employing the criterion(*14*) of $\rho_{xx}/\theta_{eff}^2$, we evaluated the energy consumption of previously reported SOT and OT channels with CoFeB or Co as the free layer as necessitated by MTJ stacks. An eight-fold reduction in energy consumption compared to Pt indicates that our Ru/W OT channel has great potential for the development of energy-efficient SOT-MRAM.

**Picosecond switching driven by OT**

After validating the giant OT generated in Ru/W OT channel, the OT-induced magnetization switching was subsequently performed on OHE-MTJ devices. We constructed a full-frequency-band probe system as depicted in Fig. 4**A** to facilitate these measurements. We employed a commercial electrical pulse generator to cover a broad temporal range from dc to 300 ps regime. Additionally, we designed and fabricated photoconductive switches (PCS) as standalone pulser modules, which, after triggered by a femtosecond laser pulse, generate ultrashort electrical pulses with a full width at half maximum (FWHM) ranging from 28.7 ps to 76 ps(*43–45*) (see Fig. 4**B** and fig. S9A for the temporal profiles of pulses). Subsequently, two picosecond pulsers were connected to 110 GHz probes (Fig. 1A), respectively. Taken the amplitude of the voltage pulse applied on the OT channel as $V_p$, a voltage pulse of $V_p/2$ is applied simultaneously on the top electrode of the MTJ for balancing the voltage drop through the MgO barrier (See Supplementary Note S5 for details). The time delay between the two electrical pulsers was precisely adjusted (Materials and Methods).

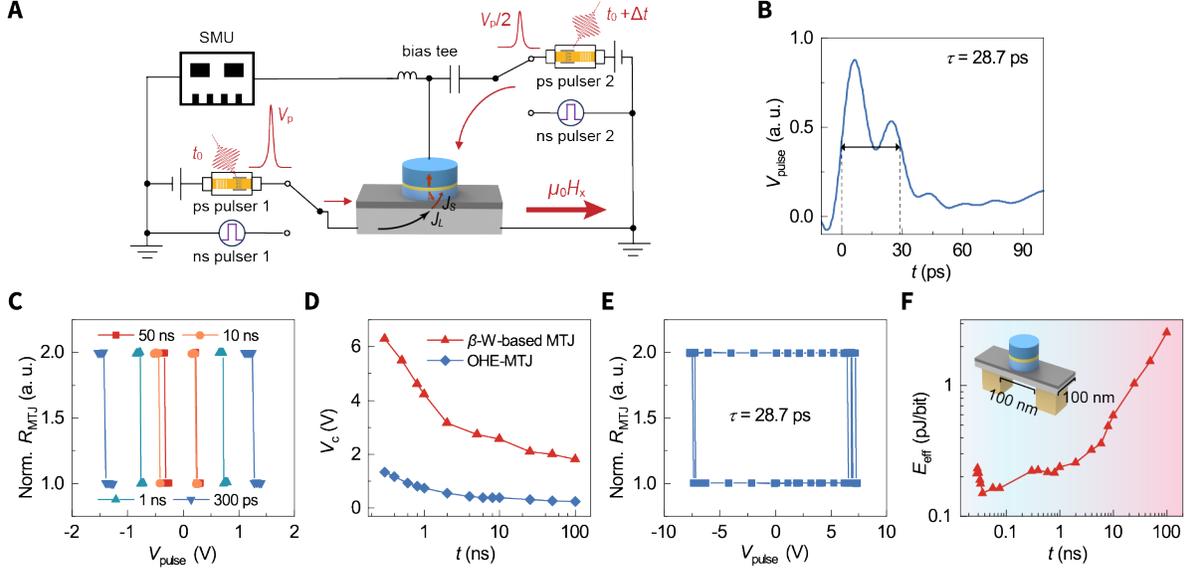

**Fig. 4. Picosecond and nanosecond switching of the OHE-MTJ.** (**A**) Schematic of the experimental set-up of our probe system. (**B**) Waveform of a picosecond pulse with a FWHM of 28.7 ps. During measurements, an external magnetic field of ±55 mT is applied for deterministic OT switching. (**C**) OT-induced switching loops within nanosecond regime. The threshold switching voltage increases with shorter pulse widths. (**D**) Comparison of threshold write voltage in the OHE-MTJ and the $\beta$-W-based MTJ versus the voltage pulse width. (**E**) OT switching of the OHE-MTJ with a speed as fast as 28.7 ps in five consecutive switching tests. (**F**) Projected energy consumption per bit of the OHE-MTJ with different pulse width. The OT channel is assumed to be a rectangular with a width and length of 100 nm (inset).

We first investigated the switching performance of the OHE-MTJs under voltage pulses ≥300 ps. A minimal voltage of 0.05 V was applied on the top electrode of the OHE-MTJs to read the TMR. Fig. 4**C** summarizes the OT-induced magnetization switching loops, with a +55 mT in-plane magnetic field applied. The switching polarity is confirmed to be the same as $\beta$-W-based MTJ. By reversing the in-plane magnetic field, an opposite switching polarity was obtained (figs. S2A and S2B), consistent with the SOT switching feature. We further calculated the critical switching current density $J_c$ for varied writing pulse widths (fig. S2C). For the pulse widths of 10 ns, 1ns and 300 ps, the obtained $J_c$ are 120, 242 and 428 MA/cm$^2$, respectively. In comparison with conventional SOT channels, the $J_c$ of OHE-MTJs is higher than $\beta$-W-based MTJs and lower than $\beta$-Ta-based MTJs. This further verifies that the $\theta_{eff}$ of Ru/W OT channel lies between $\beta$-Ta and $\beta$-W(*46, 47*). However, we note that the write voltage $V_c$ of the OHE-MTJs exhibits five to eight-fold reduction compared with that of $\beta$-W-based PMA-MTJs fabricated under identical processes, as shown in Fig. 4**D**, due to the substantially lower resistivity and, consequently, the giant $\sigma_{eff}$.

We further explore the write speed potential of our OHE-MTJs. As shown in Fig. 4**E**, we achieved OT-induced switching in our OHE-MTJs as fast as 28.7 ps with the assistance of the in-plane field of +55 mT. The robustness of the switching is confirmed through five consecutive switching events. By reversing the direction of the external field, opposite switching polarity was also obtained, as shown in fig. S2D. Assuming a rectangular OT channel with dimensions of 100 nm × 100 nm (as shown in the inset of Fig. 4**F**), we calculated the projected energy consumption $E_{eff}$ of OHE-MTJs. In nanosecond switching regime, the $E_{eff}$ of OHE-MTJs decreases with pulse width and gradually saturates at 0.3 ns. The $E_{eff}$ of 1 ns is only 237 fJ/bit, comparable to that of previously reported $\beta$-W-based MTJs(*48*). For picosecond switching, the $E_{eff}$ is calculated by integrating the pulse over a duration of 0 to 200 ps. While this approach may result in an overestimation of $E_{eff}$ (see Supplementary Note 6), the OT-induced picosecond switching demonstrates superior energy efficiency compared with nanosecond switching, in agreement with the trend reported by Díaz et al(*43*). As the pulse width decreases from 76 ps to 28.7 ps, the $E_{eff}$ shows a slight reduction before gradually increasing. The different trend of $E_{eff}$ for short picosecond pulses may originate from variations in device size, circuit design, calibration techniques or energy consumption estimation methods. Notably, a minimum $E_{eff}$ of 150 fJ/bit is obtained at a pulse width of 36.9 ps, proving that our OHE-MTJ is a highly promising for ultrafast and energy-efficient memory applications.

**OHE-MTJ for high-performance SOT-MRAM**

Featuring a giant $\sigma_{eff}$ with effective charge-to-spin conversion efficiency of -0.25 and resistivity of 20 $\mu\Omega \cdot$cm, the Ru/W OT channel shows significant merits of improving device uniformity and storage density for SOT-MRAM. For a comprehensive comparison, we analyzed the OHE-MTJs and $\beta$-W-based PMA-MTJs fabricated under the same processes. As reflected by the BE resistance collected among 72 devices fabricated on an 8-inch wafer (within a single designed die), conventional $\beta$-W-based MTJs suffer from significant distribution among devices (fig. S3A) because of the low etch-stop margin brought by the limitation of the thickness of $\beta$-W, typically below 5 nm. In contrast, the OHE-MTJs exhibit nearly uniform BE resistance with a variation of less than 1 $\Omega$. Besides, OHE-MTJs exhibit a substantial reduction in the threshold write voltage compared with $\beta$-W-based MTJs (Fig. 4D). Such ultralow switching voltages enable the integration of multiple MTJs on a single OT channel, thereby significantly reducing the area overhead of SOT-MRAM. Adopting the NAND-like architecture(*49*) depicted in fig. S3B, the number of transistors can be reduced from $2N$ to $N+1$ per SOT-MRAM device, where $N$ is the number of MTJs sharing the same write channel. Circuit-level simulations at the 28 nm node indicate that the maximum $N$ value for $\beta$-W-based MTJs is limited to 4 due to the high BE resistance (Supplementary Note S8), which prevents the voltage applied across the MTJ from achieving STT switching. In contrast, OHE-MTJs support a maximum $N$ value of 32, leading to a 37% reduction of area overhead compared with $\beta$-W-based NAND-like architecture and a 45% reduction compared with $\beta$-W-based single MTJs (fig. S3C). These results underscore the potential of OHE-MTJs as a highly promising solution for achieving high-uniformity and high-density SOT-MRAM.

**Conclusion**

In conclusion, we present a feasible approach for advancing high-performance SOT-MRAM by leveraging cutting-edge orbital effects. We demonstrate that the Ru/W OT channel, featuring a strong OHC in Ru and a conversion efficiency $C_{LS}$ exceeding 90% in $\alpha$-W, achieves a maximum $\sigma_{eff}$ of -12,631 $\hbar/2e$ $\Omega^{-1}$ cm$^{-1}$, the highest value reported to date. The Ru/W OT channel exhibits excellent compatibility with CoFeB-based MTJs, achieving a write speed as fast as 28.7 ps in Ru/W-based OHE-MTJs. Compared with conventional $\beta$-W-based SOT-MTJs, the OHE-MTJs exhibit a nearly uniform distribution of BE resistance and a five to eight-fold reduction in driving voltages. Our circuit-level simulations further indicate that up to 32 MTJs can be integrated into a single Ru/W OT channel, functioning as a high-density NAND-like SOT-MRAM array, yielding a 45% reduction in area overhead. Our work bridges the gap between orbital effects and memory applications, not only underscoring the transformative potential of the emerging field of orbitronics but also laying a robust foundation for the development of low-power, high-uniformity and high-density commercialized SOT-MRAM.

12. C. F. Pai, L. Liu, Y. Li, H. W. Tseng, D. C. Ralph, R. A. Buhrman, Spin transfer torque devices utilizing the giant spin Hall effect of tungsten. *Appl. Phys. Lett.* **101**, 122404 (2012).

13. H. Honjo, T. V. A. Nguyen, T. Watanabe, T. Nasuno, C. Zhang, T. Tanigawa, S. Miura, H. Inoue, M. Niwa, T. Yoshiduka, Y. Noguchi, M. Yasuhira, A. Tamakoshi, M. Natsui, Y. Ma, H. Koike, Y. Takahashi, K. Furuya, H. Shen, S. Fukami, H. Sato, S. Ikeda, T. Hanyu, H. Ohno, T. Endoh, "First demonstration of field-free SOT-MRAM with 0.35 ns write speed and 70 thermal stability under 400°C thermal tolerance by canted SOT structure and its advanced patterning/SOT channel technology" in *2019 IEEE International Electron Devices Meeting (IEDM)* (IEEE, 2019; https://ieeexplore.ieee.org/document/8993443/), pp. 28.5.1-28.5.4.

14. J. Han, A. Richardella, S. A. Siddiqui, J. Finley, N. Samarth, L. Liu, Room-Temperature Spin-Orbit Torque Switching Induced by a Topological Insulator. *Phys. Rev. Lett.* **119**, 077702 (2017).

15. H. Wu, A. Chen, P. Zhang, H. He, J. Nance, C. Guo, J. Sasaki, T. Shirokura, P. N. Hai, B. Fang, S. A. Razavi, K. Wong, Y. Wen, Y. Ma, G. Yu, G. P. Carman, X. Han, X. Zhang, K. L. Wang, Magnetic memory driven by topological insulators. *Nat. Commun.* **12**, 6251 (2021).

16. L. Salemi, P. M. Oppeneer, First-principles theory of intrinsic spin and orbital Hall and Nernst effects in metallic monoatomic crystals. *Phys. Rev. Mater.* **6**, 095001 (2022).

17. L. Zhu, D. C. Ralph, R. A. Buhrman, Maximizing spin-orbit torque generated by the spin Hall effect of Pt. *Appl. Phys. Rev.* **8**, 031308 (2021).

18. K. Garello, I. M. Miron, C. O. Avci, F. Freimuth, Y. Mokrousov, S. Blügel, S. Auffret, O. Boulle, G. Gaudin, P. Gambardella, Symmetry and magnitude of spin-orbit torques in ferromagnetic heterostructures. *Nat. Nanotechnol.* **8**, 587–593 (2013).

19. A. Du, D. Zhu, K. Cao, Z. Zhang, Z. Guo, K. Shi, D. Xiong, R. Xiao, W. Cai, J. Yin, S. Lu, C. Zhang, Y. Zhang, S. Luo, A. Fert, W. Zhao, Electrical manipulation and detection of antiferromagnetism in magnetic tunnel junctions. *Nat. Electron.* **6**, 425–433 (2023).

20. Y.-G. Choi, D. Jo, K.-H. Ko, D. Go, K.-H. Kim, H. G. Park, C. Kim, B.-C. Min, G.-M. Choi, H.-W. Lee, Observation of the orbital Hall effect in a light metal Ti. *Nature* **619**, 52–56 (2023).

21. G. Sala, P. Gambardella, Giant orbital Hall effect and orbital-to-spin conversion in 3d, 5d, and 4f metallic heterostructures. *Phys. Rev. Res.* **4**, 033037 (2022).

22. S. Ding, A. Ross, D. Go, L. Baldrati, Z. Ren, F. Freimuth, S. Becker, F. Kammerbauer, J. Yang, G. Jakob, Y. Mokrousov, M. Kläui, Harnessing Orbital-to-Spin Conversion of Interfacial Orbital Currents for Efficient Spin-Orbit Torques. *Phys. Rev. Lett.* **125**, 177201 (2020).

23. D. Lee, D. Go, H.-J. Park, W. Jeong, H.-W. Ko, D. Yun, D. Jo, S. Lee, G. Go, J. H. Oh, K.-J. Kim, B.-G. Park, B.-C. Min, H. C. Koo, H.-W. Lee, O. Lee, K.-J. Lee, Orbital torque in magnetic bilayers. *Nat. Commun.* **12**, 6710 (2021).
10

*The references listed below are only cited in Supplementary Materials.*

**Acknowledgments:** The author thank Xiantao Shang, Wenwen Wang, Jianing Liang, Yongxing Yang, Peng Xie and other engineers from Truth Memory corporation for the sample preparation.

**Funding:**

National Key Research and Development Program of China grant 2022YFB4400200 and 2022YFA1402604

National Natural Science Foundation of China grant T2394474, T2394472, W2411060, 52121001, 62404013 and 62401026

Beijing Outstanding Young Scientist Program

Tencent Foundation through the XPLORER PRIZE

France 2030 government investment plan managed by the French National Research Agency (ANR) under grant PEPR SPIN–[SPINTHEORY] ANR-22-EXSP-0009

EIC Pathfinder OPEN grant 101129641 "OBELIX"

**Author contributions:**

Conceptualization: YY, DZ, WZ

Methodology: XN, XG, ZY, AM, CW, ZW

Investigation: YY, CX, WC, DZ, KY, XD, SL, HZ, SC, RX, DM

Visualization: YY, ZG

Funding acquisition: WZ, DZ, AM

Project administration: WZ, DZ, HL

Supervision: WZ, HL, ZW, KC, AM

Writing – original draft: YY, DZ, CX, XN, WC, XG, DX

Writing – review & editing: YY, CX, XN, WC, XG, ZG, KY, DX, ZY, SL, HZ, SC, RX, DM, CW, ZW, DZ, KC, HL, AM, WZ

**Competing interests:** Authors declare that they have no competing interests.

**Data and materials availability:** All data are available in the main text or the supplementary materials.

**Supplementary Materials**

Materials and Methods

Supplementary Note S1-S8

Figs. S1 to S9

Table S1

References (*50-66*)

Movie S1